\documentclass{article}
\usepackage{graphicx}
\usepackage{epstopdf}
\usepackage{physics}
\usepackage{authblk}
\usepackage[margin=1in]{geometry}

\title{Two-dimensional double-quantum spectroscopy: peak shapes as a sensitive probe of carrier interactions in quantum wells}

\author[1]{\small Jonathan Tollerud}
\author[1,*]{Jeffrey A. Davis}

\affil[1]{Centre for Quantum and Optical Science, Swinburne University of Technology, 1 John St. Hawthorn, Victoria, Australia, 3122}

\affil[*]{Corresponding author: jdavis@swin.edu.au}
\date{}

\begin{document}
\maketitle

\begin{abstract}

We identify carrier scattering at densities below which it has previously been observed in semiconductor quantum wells. 
These effects are evident in the peakshapes of 2D double-quantum spectra, which change as a function of excitation density. At high excitation densities ($\geq10^{9}$\,carriers/,cm$^{-2}$) we observe untilted peaks similar to those reported in previous experiments. At low excitation densities (\textless$10^{8}$\,carriers\,cm$^{-2}$) we observe narrower, tilted peaks.
Using a simple simulation, we show that tilted peak-shapes are expected in double-quantum spectra when inhomogeneous broadening is much larger than homogeneous broadening, and that fast pure-decoherence of the double-quantum coherence can obscure this peak tilt. 
These results show that carrier interactions are important at lower densities than previously expected, and that the `natural' double-quantum peakshapes are hidden by carrier interactions at the excitation densities typically used. Furthermore, these results demonstrate that analysis of 2D peak-shapes in double-quantum spectroscopy provides an incisive tool for identifying interactions at low excitation density.
\end{abstract}


\newcommand{\ETA}[1]{$E_{#1}$}
\newcommand{\ETB}[1]{$E_{#1}$ }
\newcommand{\ETQA}{$E_{2Q}$ }
\newcommand{\ETQB}{$E_{2Q}$}
\newcommand{\EThA}{$E_{3}$ } 
\newcommand{\EThB}{$E_{3}$}
\newcommand{\FigRefA}[1]{Fig.~\ref{#1}}
\newcommand{\FigRefB}[1]{Figure~\ref{#1}}
\newcommand{\EqRefA}[1]{Eq.~\ref{#1}}
\newcommand{\EqRefB}[1]{Equation~\ref{#1}}
\newcommand{\DPW}{DP$^{WX}$ }
\newcommand{\DPN}{DP$^{NX}$ }


\section{Introduction}
\small
The coherent dynamics of excitons 
in semiconductor quantum wells (QWs) are highly dependent on carrier-carrier interactions ~\cite{Chemla2001,Cundiff2008} 
One consequence of the strongly interacting excitons is the formation of multiple exciton correlations~\cite{Bolton2000,Stone2009, Turner2009}.
The simplest such correlation is a two-exciton correlation, which loosely speaking corresponds to the correlated motion of two electons and two holes. 
Two types of two-exciton correlations have been observed: those which are bound through Coulomb attraction (usually called biexcitons), and those that are made up of unbound exciton pairs.
Direct optical excitation of biexcitons or unbound two-exciton correlations from the ground state is optically forbidden, so they must be excited and measured as multiple photon processes. As a result the signal from two exciton correlations is often difficult to disentangle from the typically much more intense single exciton signals.

Biexcitons often can be separated spectrally from single exciton emission: 
the biexciton binding energy (on the order of 1-2\,meV in GaAs~\cite{Stone2009,Miller1982}) red-shifts the biexciton to exciton transition energy. thereby allowing it to be distinguished from the single exciton transition. 
Unlike biexcitons, signatures of unbound two exciton correlations are not usually shifted from the one-exciton emission, and therefore cannot be easily separated spectrally.

\begin{figure}[t!]
\centering
\fbox{\includegraphics[scale=1]{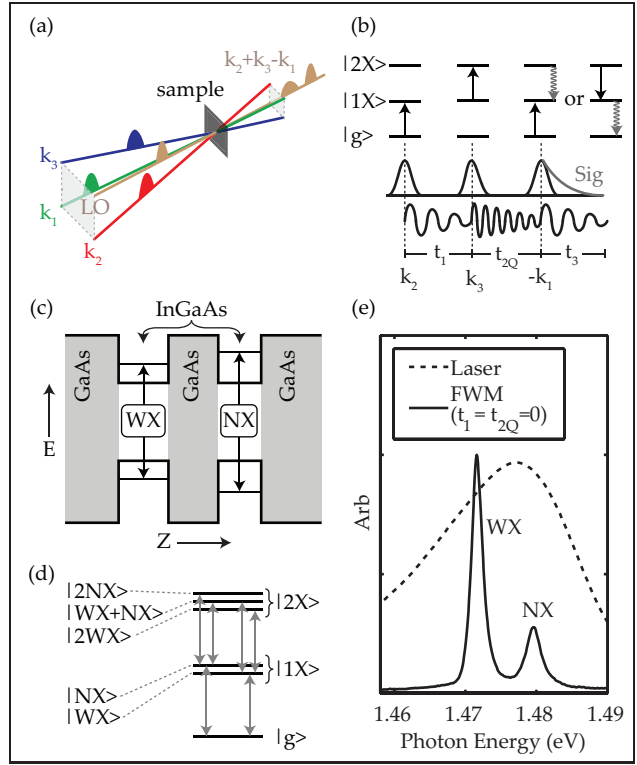}}
\caption{(a) The box beam geometry used in the 2Q 2D experiment. (b) The pulse ordering in the 2Q experiment: the conjugate pulse ($k_1$) arrives last. $\ket{1X}$ and $\ket{2X}$ represent singly and doubly excited states, respectively. The phase of the signal (shown below the pulses) varies at the optical frequency during $t_1$ (where the system is in a one-quantum coherent superposition of $\ket{g}$ and $\ket{1X}$), and then at twice the optical frequency during $t_{2Q}$ (where the system is in a two-quantum coherent superposition of $\ket{g}$ and $\ket{2X}$), the third pulse returns the system to a single quantum coherence, which then emits the signal. (c) The DQW structure, which is made up of InGaAs QWs sandwiched between GaAs barriers. We consider here only the lowest energy exciton transitions in each well, which are labelled WX and NX. (d) The energy levels of the DQW considering correlations of up to two excitons. The states are separated into a one-exciton ($\ket{1X}$) manifold with two levels ($\ket{WX}$ and $\ket{NX}$) and a two-exciton ($\ket{2X}$) manifold with three levels ($\ket{2WX}$, $\ket{2NX}$ and $\ket{WX+NX}$). (e) The excitation spectrum used in the 2Q experiments and the emission spectrum of the FWM signal at pulse overlap. (color online)}
\label{Fig:DQW}
\end{figure}

Two-exciton correlations can, however, be separated from the one-exciton signals in double-quantum (2Q) 2D electronic spectroscopy. In 2Q spectroscopy, three mutually coherent fs or ps pulses (labelled $k_2$, $k_3$ and $k_1$) in the box geometry shown in \FigRefA{Fig:DQW} (a) are used to generate a four-wave mixing (FWM) signal, which is emitted in the phase matched direction $k_{FWM} = k_2+k_3-k_1$. The signal is combined with a co-propogating reference pulse (called the local oscillator or LO) to generate a spectral interferogram, which provides access to the spectral phase and amplitude of the signal electric field. 
The order of the pulses is such that $k_1$ (the conjugate pulse) arrives last, as shown in \FigRefA{Fig:DQW} (b). 
The first two pulses generate a 2Q-coherence between $\ket{2X}$ and $\ket{g}$ in a two step process. The phase of the 2Q-coherence then oscillates at roughly twice the optical frequency of $\ket{1X}$ where the frequency of $\ket{1X}$ is comparable to the frequency of the laser electric field. This 2Q-coherence is non-radiative, so it cannot be directly read out. Instead, the final excitation pulse returns the system to a one-quantum coherence, which then radiates in the form of the FWM signal. 

The spectral amplitude and phase of the FWM signal are recorded as a function of the delay between the second and third pulse (labelled $t_{2Q}$) for a constant delay between the first two pulses (labelled $t_1$). A Fourier transform is then applied to the recorded data as a function of $t_{2Q}$ to generate 2D 2Q spectra, in which signals are spread along the {\ETQB}. Signals are thus separated according to both their emission energy and the energy of the 2Q-coherence (i.e. the frequency of the phase oscillations) during $t_{2Q}$.

Importantly, signals that do not involve a 2Q-coherence during $t_{2Q}$ do not generate signal in the detection direction. As a result, all of the signal in the 2Q 2D spectrum results from some type of 2Q-coherence (e.g. two-exciton correlations), and signals involving only single exciton states are suppressed. Furthermore, the details of the 2Q-coherence can be discerned from the position of the peak in the 2Q 2D spectrum. A two-exciton correlation for excitons with transition energy $\epsilon_A$ and $\epsilon_B$ will generate a signal at $E_{2Q}=\epsilon_A+\epsilon_B$. Biexcitons of A and B appear at $E_{2Q}=\epsilon_A+\epsilon_B-\epsilon_{BX}$ where $\epsilon_{BX}$ is the biexciton binding energy.

Previous 2Q experiments on QWs have shown that signals can be generated from bound-biexcitons or from unbound pairs of exctions~\cite{Stone2009,Stone2009b,Turner2009,Karaiskaj2010}.
The coherence time of biexcitons have been measured, and found to depend on excitation density~\cite{Stone2009,Stone2009b}.
The expected polarization selection rules of mixed and non-mixed biexcitons (based on the magnetic quantum number of the hole states) have been confirmed~\cite{Stone2009,Bristow2009}.

While the stable state of biexcitons can be readily understood (they involve the binding of excitons with opposite spin), the details of the mechanism underpinning the unbound exciton correlations are less clear. The interactions could be considered as a two-body (four-body) interaction between the two excitons (two electrons and two holes) mediated, for example, by Coulomb forces. Alternatively the signal from the unbound two-exciton state could be mediated by a many-body coupling involving mean field interactions between the two states (as was proposed in Ref.~\cite{Karaiskaj2010}). 

Many of the results and conclusions drawn from these previous experiments rely (at least in part) on analysis of peak-shapes, so a detailed understanding of the factors that affect the 2Q peak-shapes must be considered. 
In particular, given that exciton dynamics are sensitive to excitation density~\cite{SHAH1999, Cundiff2008, Stone2009, Stone2009b}, the dependence of the 2Q peak-shapes on the photon density of the excitation pulses merits investigation.

In this article, we study the coherent dynamics of correlated exciton pairs through 2Q spectroscopy, focussing on the dependence of the 2D peak-shapes and 2Q linewidths on the photon density of the excitation pulses. We perform measurements down to photon densities of $4\times 10^{8}$\,photons\,cm$^{-2}$ - much lower than previous 2Q 2D spectroscopy experiments - and find that the 2D spectra change in significant ways. 
At the lowest photon densities, we observe 2D 2Q peak-shapes that are tilted towards the $E_{2Q} = 2E_3$ line and \ETQA linewidths that are limited by the inhomogeneous linewidths of the two correlated exciton states. 
This is in contrast to previous 2Q experiments on QWs, in which untilted 2Q peak-shapes and larger $E_{2Q}$ linewidths are typically reported~\cite{Stone2009,Stone2009b,Turner2009,Karaiskaj2010}.
At intermediate photon densities (still much lower than previous experiments) we find that the 2Q linewidth increases monotonically with photon density, which we take to indicate that the linewidths are limited by exciton-carrier (exciton-free carrier and/or exciton-exciton) scattering.
At excitation densities comparable with those used in previous experiments, we observe peak-shapes which are not tilted and have enhanced 2Q linewidths.


\section{Experiment}
\subsection{Double-quantum well sample}

The sample studied here is a coupled asymmetric double-quantum well (DQW), which is a layered semiconductor heterostructure grown by molecular beam epitaxy. A cartoon diagram of the DQW is shown in \FigRefB{Fig:DQW}\,(c). It consists of two layers of In$_{0.05}$Ga$_{0.95}$As (the QWs) sandwiched between layers of GaAs (the barriers). The wide well is 10\,nm thick, the narrow well is 8\,nm thick and the central barrier is 10\,nm thick. Each QW has one `bright' exciton transition (an n=1 electron state in the conduction band with an n=1 heavy-hole state in the valence band), labelled WX and NX for the wide well and narrow well respectively. The light-hole valence band levels in this structure are not confined in the QWs due to strain~\cite{Marzin1985,Pan1988}, so there are no other exciton transitions in this spectral range. As shown in \FigRefB{Fig:DQW}\,(d), when we consider correlations of up to two excitons, there are a total of six states: one ground state ($\ket{g}$), two singly excited states ($\ket{WX}$ and $\ket{NX}$) and three doubly excited states ($\ket{2WX}$, $\ket{2NX}$ and $\ket{WX+NX}$). It is convenient to separate the excited states into a single exciton manifold ($\ket{1X}$) and a two exciton manifold ($\ket{2X}$). Direct transitions between $\ket{g}$ and $\ket{2X}$ are optically forbidden.

Although the outer barriers confine the carriers within the DQW structure, the relatively low and narrow central barrier allows some penetration of the electron wavefunctions from one well into the other. Wavefunction calculations (shown in Ref.~\cite{Nardin2014}) show hybridization of the electron wavefunctions across the entire DQW structure. Furthermore, experiments in the same reference demonstrated that the WX and NX can interact through the appearance of cross-peaks in 2Q and 1Q spectra.
Using this coupled DQW structure we can compare the dynamics of two-exciton correlations in which the excitons are localized in either the same QW or in different QWs.


\subsection{2Q spectroscopy}

We perform 2Q 2D spectroscopy using a diffraction based pulse shaper as was first demonstrated by Turner et al.~\cite{Turner2011}. 
The inter-pulse delays are controlled through the application of linear spectral phase gradients to each of the beams independently using the pulse shaper. 
The delays are applied in a rotating frame by fixing the phase for a particular spectral component (1.459\,eV in this case), which reduces the sampling requirements compared with experiments using non-rotating frame.
Delaying the pulses in a rotating frame is particularly useful in 2Q spectroscopy, because the oscillations of the signal phase as a function of t$_{2Q}$ occur at twice the optical frequency, which corresponds to a period of ${\sim}$1\,fs at 850\,nm. The rotating frame allows us to use ${\sim}$10\,fs steps while still oversampling in t$_{2Q}$.

Another benefit of this approach is that all of the beams are incident upon the same optics, so the beams are intrinsically phase stabilized relative to each other. This stability allows us to make measurements at much lower photon densities than are typically used in 2Q spectroscopy.
The beams are focussed to a 150\,$\mu$m spot and overlapped at the sample (which is held in a vibration isolated cryostat at 6\,K) using an f\,=\,12.5\,cm spherical lens.

The photon density was varied from $4.3\times 10^{8}$\,cm$^{-2}$ to $2\times 10^{11}$\,cm$^{-2}$ using neutral density filters. Although we did not measure the absolute absorption coefficient for this sample, we can estimate the excitation density by assuming a typical absorption coefficient of 1-4\% per well~\cite{Masselink1985}. At the lowest photon densities presented here, that results in an excitation density of ${\sim}10^{7}$\,excitons\,cm$^{-2}$.

The beams were all co-circularly polarized to suppress signals involving bound biexcitons. 
The excitation spectrum and the FWM signal at $t_1 = t_{2Q} = 0$\,fs can be seen in \FigRefB{Fig:DQW}\,(e). 
The 2Q spectra are collected by scanning $t_{2Q}$ from 0 to 2\,ps in 10\,fs steps for $t_1=0$\,fs, and are plotted as the amplitude of the signal electric field. 
A full description of our experimental apparatus can be found in Ref.~\cite{Tollerud2014}.

\section{Results and discussion}

\begin{figure}[t!]
\centering
\fbox{\includegraphics[width=0.6\linewidth]{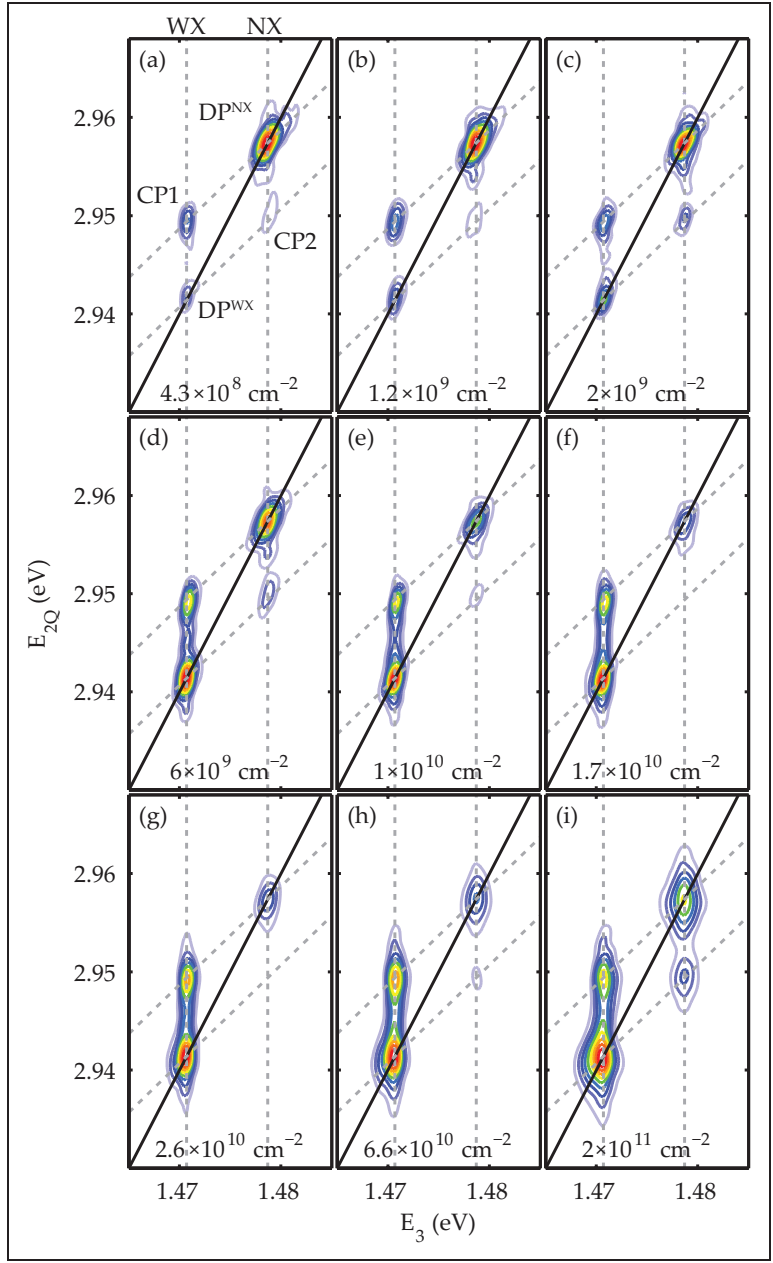}}
\caption{2Q 2D absolute value spectra of the InGaAs DQW for different photon densities (which are given in the bottom of each spectrum). The spectra are all self-normalized and plotted on a linear colour scale with contours drawn at 10\% intervals. (color online) }
\label{Fig:Data2D}
\end{figure}

2Q 2D amplitude spectra of the DQW are shown in \FigRefA{Fig:Data2D} for a range of different photon densities. The spectrum of the emitted signal (\EThB) is shown along the horizontal axis and the the corresponding \ETQA energy can be seen along the vertical axis. The diagonal solid line indicates the \ETQB\,=\,2\,\EThA line. 
Signals in which both of the correlated excitons are of the same type (i.e. the photons from the first two pulses interacted with the same exciton transition) can be found along this line.
Signals in which the both excitons that make up the two-exciton state are from the same transition can be found along this line.  
We observe two such diagonal-peaks (DPs), one for WX and one for NX, which are labelled \DPW and \DPN respectively. 
We also observe two cross-peaks (CPs), one at \EThB\,=\,1.471\,eV (WX) and one at \EThB\, = \,1.478\,eV (NX), which are labelled CP1 and CP2, respectively. 
Both CPs appear at \ETQB\,=\,1.471\,+\,1.478\,=\,2.549\,eV (WX\,+\,NX) which is where we expect to see 2Q coherences involving a mixed two exciton state made up of one WX exciton and one NX exciton. 

The shapes of both the DPs and CPs change as a function of the photon density.
At low photon density, we observe tilted peak-shapes, in which the peaks have a major axis aligned roughly along the \ETQB\,=\,2\,\EThA line. 
As the photon density increases, the peaks broaden in the \ETQA direction, and the tilt of the peaks becomes less evident.
At the highest photon densities, we observe virtually no tilt of the CPs or DPs. 

The relative amplitude of the peaks also changes significantly as a function of photon density. 
An analysis of the dependence of the absolute amplitude of the DPs and CPs on photon density will be the subject of a future publication, but will not be considered here.


\subsection{Tilted 2Q peak-shapes}

In the more commonly used 1-Quantum (1Q) 2D spectroscopy (in which the conjugate pulse arrives first and the time between the first and second pulse is scanned), tilted DPs are indicative of inhomogeneous broadening, and tilted CPs are indicative of energetically separate transitions in which the inhomogeneous broadening is correlated. The cross-diagonal width of the DP is related to the homogeneous linewidth, while the width along the diagonal is related to the inhomogeneous linewidth. 

In contrast to 1Q spectroscopy, the origin of tilted 2Q peaks is an inherent correlation of \ETQA and \EThB. Qualitatively, this can be understood as follows: \EThA for a particular pathway within the inhomogeneous ensemble is determined by the energy of the 2Q coherence (\ETQB) and the energy at which the third pulse interacts with the 2Q coherence ({$E_{k1}$) as \EThB\,=\,\ETQB\,-\,\ETA{k1}. As a result, \EThA and \ETQA are intrinsically correlated, which manifests experimentally as a tilt of the peaks. Importantly, this tilt does not originate from correlations in the sample, but rather is a consequence of the measurement technique itself. 

Tilted 2Q peaks have been previously observed in some molecular systems~\cite{Kim2009, Nemeth2010, Christensson2010}, but have (to our knowledge) never been reported in measurements on QWs. Previous measurements on QWs with 2Q spectroscopy have exhibited peaks that are not-tilted and are more than twice as broad along \ETQA than along \EThB~\cite{Karaiskaj2010,Stone2009,Turner2009}. The combination of a lack of tilt and an additional broadening along \ETQA suggests that there is some additional decoherence of the $\ket{g}$-$\ket{2X}$ coherence that does not affect the individual $\ket{g}$-$\ket{1X}$ coherences, thus increasing the width of the peak along \ETQA and obscuring the tilt.

To test this explanation for the untilted peak-shapes, we simulate 2Q 2D peaks. We first solve the optical Bloch equations for a single pathway in a two-level system, assuming delta function pulses, $t_1 = 0$\,fs and (for now) ignoring inhomogeneous broadening. Under these assumptions, the third order polarization P$^{(3)}$ (which is proportional to the measured signal in an interferometrically detected 2Q 2D spectroscopy experiment) is given by:

\begin{equation}\label{Eq:Pol1}
P^{(3)}(t_{2Q},t_3) = \textup{ exp}(i\omega_{2Q}t_{2Q}+i\omega_3t_3 - \gamma_{2Q}t_{2Q} - \gamma_3t_3)
\end{equation}
where $\gamma_{2Q}$ is the decoherence rate of the $\ket{g}$-$\ket{2X}$ 2Q-coherence in $t_{2Q}$ and $\gamma_3$ is the decoherence rate of the $\ket{g}$-$\ket{1X}$ 1Q-coherence in $t_3$. $\gamma_{2Q}$ can be defined based on the decoherence rates of the transitions with which the first two pulses are resonant ($\gamma_{k2}$ and $\gamma_{k3}$) and an additional term ($\tilde{\gamma}_{2Q}$) which represents decoherence inducing interactions that affect the $\ket{g}$-$\ket{2X}$ coherences but not the $\ket{g}$-$\ket{1X}$ coherences:
\begin{align}\label{Eq:Dec1}
\gamma_{2Q} &= \gamma_{k2}+\gamma_{k3}+\tilde{\gamma}_{2Q}
\end{align}
As described above, $\omega_{2Q}$ = $\omega_{k1}+\omega_3$, so \EqRefA{Eq:Pol1} can be rewritten as:
\begin{equation}
P^{(3)}(t_{2Q},t_3) = \textup{ exp}\left[i(\omega_{k1}+\omega_3)t_{2Q}+i\omega_3t_3+R(t_{2Q},t_3)\right] 
\end{equation}
where R is a relaxation term which includes all of the decoherence rates:
\begin{equation}\label{Eq:Pol2}
R(t_{2Q},t_3)=-(\gamma_{k2}+\gamma_{k3}+\tilde{\gamma}_{2Q})t_{2Q} - \gamma_3t_3
\end{equation}
At 6\,K, and with the excitation densities used here, the inhomogeneous linewidth of QW excitons is typically much larger than the homogeneous linewidth~\cite{SHAH1999, Cundiff2008}. The peak-shapes in 2Q 2D spectra are therefore mostly determined by inhomogeneous contributions. The inhomogeneous broadening of the $\ket{2X}$ states depends on the inhomogeneous broadening of each of the associated $\ket{1X}$ states, which can be introduced by integrating across Gaussian distributions of energies for the excitons resonant with $\omega_{k1}$ and $\omega_{3}$:
\begin{equation}\label{Eq:Pol22}
\begin{split}
P^{(3,IH)}(t_{2Q}, t_3)=&\int_0^{\infty}\partial \omega_{3}\int_0^{\infty}\partial \omega_{k1}\\
&G_{\mu_{k1},\Delta_{k1}}(\omega_{k1})G_{\mu_{3},\Delta_{3}}(\omega_{3})P^{(3)}(t_{2Q},t_3)
\end{split}
\end{equation}
where $G_{\mu,\Delta}(\omega) = \textup{exp}\left[-(w-\mu)^2/2\Delta^2\right]$ is a standard Gaussian distribution. $\mu_3$ ($\Delta_3$) is the center (width) of the inhomogeneous distribution of the FWM signal and $\mu_{k1}$ ($\Delta_{k1}$) is the center (width) of the inhomogeneous distribution of the transition with which the third pulse is resonant. After integration, \EqRefA{Eq:Pol22} becomes:

\begin{equation}\label{Eq:Pol3}
\begin{split}
P^{(3,I\!H)}&(t_{2Q}, t_3)=\\
&\textup{exp}\left[i\mu_3t_3+i(\mu_{k1}+\mu_3)t_{2Q}+R(t_{2Q},t_3)  \vphantom{\textup{exp}-\frac{1}{2}(t_{2Q}^2(\Delta_{k1}^2+\Delta_3^2)+2(t_{2Q}+t_3)\Delta_3^2+t_3^2\Delta_3^2)}\right. \\
&\left.-\frac{1}{2}(t_{2Q}^2(\Delta_{k1}^2+\Delta_3^2)+2 t_{2Q} t_3\Delta_3^2+t_3^2\Delta_3^2)\right]
\end{split}
\end{equation}

The third line in the exponential in \EqRefA{Eq:Pol3} is the contribution from the inhomogeneous broadening. The correlation of \ETQA and \EThA appears due to the cross term including both $t_3$ and $t_{2Q}$. If the $\Delta$ terms are much larger than the $\gamma$ terms (which is to say, that the inhomogeneous width is much larger than than the homogeneous width), then the inhomogeneous term will predominantly determine the peak-shape, which will thus be tilted. If any of the terms within R are comparable with the inhomogeneous widths, then the homogeneous broadening will also contribute meaningfully to the peak-shape, reducing or obscuring the tilt. Since the homogeneous broadening is much smaller than the inhomogeneous broadening for these photon densities and at this sample temperature, we expect to see peaks which are tilted.

By convention, in a rephasing four-wave mixing experiment, the delay between the conjugate pulse and the first non-conjugate pulse is taken to be positive when the conjugate pulse arrives first. As a result, $t_{2Q}$ is taken to be negative here because the conjugate pulse arrives after the non-conjugate pulses. 

\begin{table}[t]
\centering
\caption{Parameters used in simulated 2Q spectra.}
\begin{tabular}{cccccccc}
\hline
 & $\Delta_{k1}$ & $\Delta_{3}$ & $\mu_{k1}$ & $\mu_{3}$ & A & $\gamma_{k2}$, $\gamma_{k3}$ & $\gamma_{3}$  \\
 & meV & meV & eV & eV & arb & meV & meV \\
\hline
\DPW & 3.5 & 3.5 & 1.471 & 1.471 &1 & 0.05, 0.05 & 0.05  \\
\DPN & 4 & 4 & 1.478 & 1.478 &0.7 & 0.05, 0.05 & 0.05 \\
CP1  & 3.5 & 4 & 1.471 & 1.478 &0.6 & 0.05, 0.05 & 0.05 \\
CP2  & 4 & 3.5 & 1.478 & 1.471 &0.4 & 0.05, 0.05 & 0.05 \\
\hline
\end{tabular}
  \label{Tab:Params}
\end{table}
\begin{figure}[t]
\centering
\fbox{\includegraphics[width=0.6\linewidth]{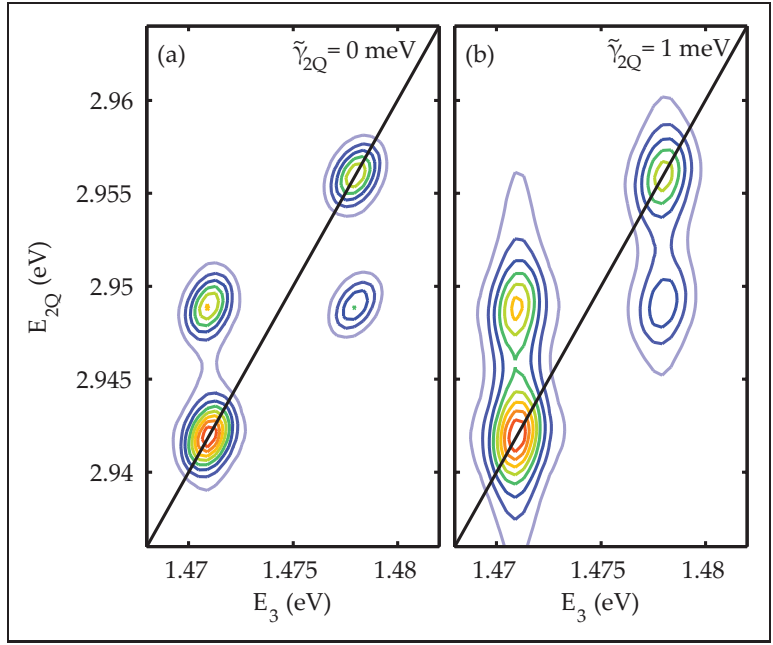}}
\caption{Simulated spectra with (a) $\tilde{\gamma}_{2Q}$\,=0 and (b) $\tilde{\gamma}_{2Q}$\,=1\,meV. Each peak was individually calculated using \EqRefA{Eq:Pol3} and combined to simulate the 2D spectrum. The peaks in (a) have a tilt towards the 2:1 line, comparable to the tilt seen in the experimental results at low photon density. The tilt is obscured in (b), resulting in a peak shapes that are comparable to our experimental results at high photon density. (color online)}
\label{Fig:SimPeaks}
\end{figure}

To generate simulated spectra, each of the four peaks (DP$^{WX}$, DP$^{NX}$, CP1 and CP2) are individually calculated using \EqRefA{Eq:Pol3}, then combined after a 2D Fourier transform. Simulated spectra are calculated for two different values of $\tilde{\gamma}_{2Q}$, which are shown in \FigRefA{Fig:SimPeaks}. When $\tilde{\gamma}_{2Q}$\,=0, the resulting peaks are all tilted towards the 2:1 line, as shown in \FigRefA{Fig:SimPeaks} (a). These spectra are qualitatively comparable to the experimental results at low photon density in \FigRefA{Fig:Data2D} (a)-(e). In contrast, when $\tilde{\gamma}_{2Q}$\,=1\,meV, the tilt is obscured and the spectra look similar to \FigRefA{Fig:Data2D} (h)-(i).
The rest of the parameters used in these simulations were chosen to qualitatively match the experimental peaks and are given in Table~\ref{Tab:Params}. 

The purpose of these simulations is not a quantitative comparison with our results, but rather a means to illustrate that the tilted peaks are not unexpected in 2Q 2D spectra, and that a phenomenological decoherence term that affects the 2Q state but not the single quantum coherence can obscure this tilt and produce peaks comparable in shape to those reported in previous 2Q 2D spectra of QW excitons.

\subsection{$E_{2Q}$ and $E_3$ and linewidths}

These simulated peaks also show that the linewidth along $E_{2Q}$ should increase as the tilt disappears with increasing photon density. Qualitatively considering the peaks in \FigRefA{Fig:Data2D}, it is clear that both the \ETQA and \EThA linewidths of all of the peaks change as a function of photon density. We now quantify this photon density dependent linewidths by fitting the peaks. $\Gamma_3^P$ (the emission linewidth of peak `P') is characterized by projecting the peak onto the \EThA axis, and then fitting each peak to a Lorentzian distribution. 

To quantify the linewidths of the peaks along \ETQB, we perform a fit along along \ETQA for each \EThA emission energy.
To account for the overlap of the peaks along {\ETQB}, each emission energy is fit to the sum of two Lorentzian peaks - one representing the DP and one representing the CP. $\Gamma_{2Q}^P$ (the average linewidth of peak `P' along the $E_{2Q}$ direction) is then calculated by averaging the $\Gamma_{2Q}$'s extracted from the 1D fits performed at each E$_3$.

\begin{figure}[t]
\centering
\fbox{\includegraphics[width=0.6\linewidth]{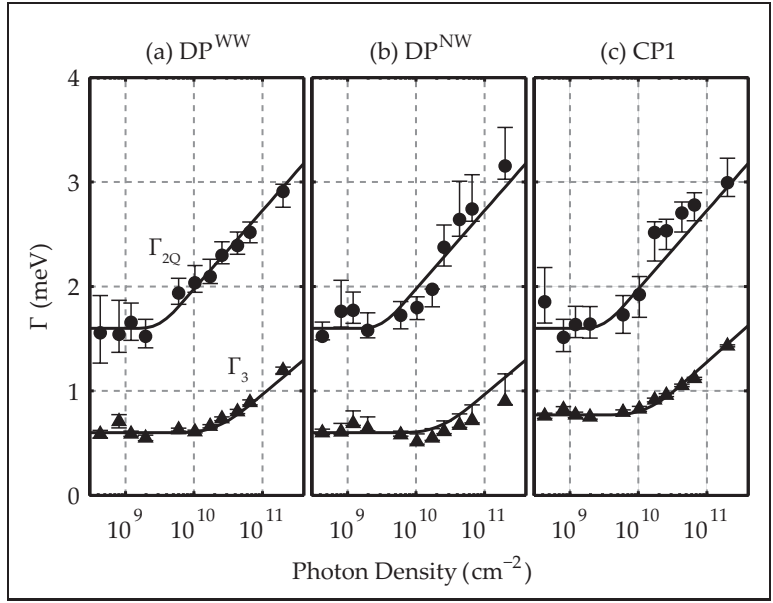}}
\caption{Peak-widths extracted from fits to the peaks in \FigRefA{Fig:Data2D} for the DPs and CP1.
The error bars are the 95\% confidence interval, and the solid lines are a guide to the eye. The $\Gamma_{2Q}$ guide lines are the same for all three peaks, while the $\Gamma_3$ lines are the same for \DPW and DP$^{NX}$, but shifted slightly for CP1. }
\label{Fig:Data1D}
\end{figure}

$\Gamma_{2Q}^P$ (filled circles) and $\Gamma_{3}^P$ (filled triangles) extracted from fits to \DPW, \DPN and CP1 are shown in \FigRefA{Fig:Data1D}. The error bars represent the 95\% confidence interval of the fit. 
Two regimes can be observed: At low density the widths do not change with photon density, while at higher photon density the peak widths increase monotonically, with a dependence that appears to be roughly logarithmic and with an identical slope. 
The solid lines are a guide to the eye to illustrate these two regimes and at which photon density the transition occurs. In all cases, the $\Gamma_{2Q}$  transition occurs at roughly $2\times10^{9}$\,cm$^{-2}$, while the $\Gamma_{3}$ transition occurs at roughly $1\times10^{10}$\,cm$^{-2}$. 

\subsection{Discussion}

The dependence of both the peak-shapes and the E$_{2Q}$ linewidth on the photon density of the excitation pulses used in the experiment are consistent with a decoherence of the $\ket{2X}$ state that is dependent on photon density (and thus on excitation density). 
In principle there are several different factors that contribute to the linewidth of the 2-exciton correlations: exciton-carrier scattering (including exciton-free carrier and exciton-exciton scattering), exciton-phonon scattering, inhomogeneous broadening and decoherence induced by disorder and defects~\cite{SHAH1999}. 
Of these possible sources of decoherence, only exciton-carrier scattering scales with excitation density.
Therefore, the monotonic increase of the linewidth above photon densities of ${\sim}2\times 10^9$\,cm$^{-2}$ suggests that the limiting factor in the decoherence of the $\ket{2X}$ linewidths is exciton-carrier scattering, and not the inhomogeneous linewidth. 
Below ${\sim} 2\times 10^9$\,cm$^{-2}$, the constant linewidth likely indicates that at these low excitation densities we are limited by one of the other decoherence mechanisms which is not dependent on excitation density. 
The fixed $\ket{2X}$ linewidths below ${\sim} 2\times 10^9$\,cm$^{-2}$, are approximately twice the inhomogeneously broadened $\ket{1X}$ linewidths. The $\ket{2X}$ linewidths are therefore likely limited by inhomogeneous broadening at the lowest photon densities.

This result is not entirely surprising, in that it is well known that the homogeneous linewidth of excitons is broadened by exciton-carrier scattering and thus depends on the excitation density~\cite{SHAH1999, Cundiff2008}. Furthermore, previous experiments have shown that the decoherence rates of biexcitions depend on excitation density as well~\cite{Stone2009,Stone2009b}.
However, the fact that we must include an additional decoherence term which affects only $\ket{g}$-$\ket{2X}$ coherences in order to reproduce the experimental spectra and that the appearance of excitation induced decoherence of 2Q-coherences occurs at photon densities an order of magnitude lower than for the $\ket{g}$-$\ket{1X}$ coherences merits additional consideration. 

Three different sources could contribute to the larger than expected dependence of the $\ket{g}$-$\ket{2X}$ decoherence times on photon density:
\begin{enumerate}
\item The scattering rate is greater for the $\ket{2X}$ coherence than for the two $\ket{1X}$ coherences
\item The decoherence induced by each elastic scattering event is larger for the $\ket{2X}$ than for the two $\ket{1X}$ coherences
\item There are additional decoherence mechanisms which affect the $\ket{2X}$ but not the $\ket{1X}$ coherences.
\end{enumerate}
While more experiments are needed to determine the source of this additional excitation induced decoherence, possibilities 1 and 2 both provide plausible explanations and imply that the 2Q peak shape is a sensitive probe of elastic interactions and many-body effects.
An increased scattering rate for the $\ket{2X}$ states (possibility 1) is plausible because the spatial cross-section of the two-exciton correlation is much larger than for the one-exciton state, as the two correlated excitons need not be spatially overlapped. 
Possibility 2 is also plausible because the size of the phase shift induced by each decohering interaction is larger for a two-exciton correlation: a scattering event that causes a phase delay of 0.18\,fs leads to a ${\sim}\frac{\pi}{8}$ phase shift to the $\ket{g}$-$\ket{1X}$ coherences, but a 2$\times$ larger ${\sim}\frac{\pi}{4}$ phase shift to the $\ket{g}$-$\ket{2X}$ coherences. Aggregated across the ensemble of coherent superpositions, this will lead to a 2$\times$ reduction of a macroscopic $\ket{g}$-$\ket{2X}$ coherence compared with a macroscopic $\ket{g}$-$\ket{1X}$ coherence for the same number of scattering events. 
It should be noted that both of these possible sources of decoherence are in addition to the ${\sim}$2$\times$ increase in the $\ket{g}$-$\ket{2X}$ decoherence rate due to the fact that either of the two correlated excitons can interact with other carriers (i.e. even if $\tilde{\gamma}_{2Q}=0$, $\gamma_{2Q}$ calculated from \EqRefA{Eq:Dec1} is still the sum of two $\ket{1X}$ decoherence rates).   

This dependence of the linewidth and peak-shape on photon density is evident and quantitatively very similar not only for \DPW and DP$^{NX}$, but also for CP1.
The broadening of CP1 along \ETQA due to exciton-carrier scattering appears at the same photon density as \DPW and DP$^{NX}$, and the rate of increase with photon density is almost identical. The similar behaviour of these peaks suggests that the decoherence rate of the $\ket{WX+NX}$ state tracks roughly with the decoherence rates of $\ket{2WX}$ and $\ket{2NX}$. Intuitively, we might expect $\ket{WX+NX}$ to decohere more rapidly for the same excitation density since it involves correlations of excitons localized predominantly in separate wells, and therefore has the opportunity to scatter with carriers in either well and thus experiences a larger effective carrier density. The similarity of the DP and CP decoherence rates and peak-shapes can be explained if we consider that the QWs are coupled and there is significant hybridization of the electron wavefunctions across the entire DQW. Excitons are therefore able to interact with carriers in the other QW, meaning that carriers in the wide well can induce decoherence of excitons in the narrow well (and vice versa). As a result, the decoherence rates of all of the correlated states in the $\ket{2X}$ manifold are comparable because they experience similar effective carrier densities.

\section{Conclusion}

In summary, we have performed 2Q 2D spectroscopy on a semiconductor DQW, using photon densities which span almost three orders of magnitude, focussing on excitation densities which are lower than typically used in 2Q spectroscopy of QWs. We observe a change in peak-shape, and the linewidth along $E_{2Q}$ as a function the photon density. In spectra collected using photon densities at the low end of the measured range (below ${\sim} 2\times 10^{10}$\,cm$^{-2}$), we observe tilted peak-shapes (for both DPs and CPs) and widths along $E_{2Q}$ which approach the sum of the linewidths of the two correlated transitions. At the higher end of the measured range, we observe un-tilted peaks, and $\ket{2X}$ linewidths which are much larger than the sum of the associated $\ket{1X}$ linewidths. The linewidths of the two-exciton correlations can be separated in two regimes, the transition between which occurs at ${\sim} 2\times 10^9$\,cm$^{-2}$ for $\ket{g}$-$\ket{2X}$ coherences and at densities almost an order of magnitude higher for $\ket{g}$-$\ket{1X}$ coherences. Our results show that at photon densities above this transition, the linewidth is limited by exciton-carrier scattering, while at photon densities below this transition the linewidth appears to be limited mostly by inhomogeneous broadening. 

These results suggest that the untilted peak-shapes which are typically reported for QW excitons are likely a result of rapid decoherence of the 2Q coherence due to exciton-carrier interactions and therefore that the measured peak-shapes and linewidths depend heavily on the excitation density. 
Furthermore, these results demonstrate that 2Q peakshape analysis is a sensitive probe of carrier-carrier scattering in QWs at low excitation densities.


\section*{Funding Information}

This work was supported by the Australian Research Council (FT120100587).

\section*{Acknowledgments}

The authors thank S. Cundiff, G. Nardin and F. Morier-Genoud for providing the DQW sample, and H. Li for helpful conversations.

\bibliographystyle{plain}
\bibliography{MyPapersTwoQLineShape}

\end{document}